\documentclass[%
preprint,
 amsmath,amssymb,
aps,
]{revtex4-1}

\usepackage{graphicx}
\usepackage{dcolumn}
\usepackage{bm}
\usepackage{color}
\usepackage{ulem}

\usepackage{txfonts}

\newcommand{\Tc}{T_{\rm c}}

\newcommand{\LaTAl}{La$\it Tr\rm_{2}$Al$_{20}$}

\newcommand{\RxVAl}{$\it R_{x}$V$_{2}$Al$_{20}$}

\begin{document}

\preprint{APS/123-QED}

\title{Superconductivity enhanced by $d$-band filling in La$\it Tr\rm_{2}$Al$_{20}$ with $\it Tr$ = Mo and W}


\author{Rumika Miyawaki}
\author{Naoki Nakamura, Ryuji Higashinaka, Tatsuma D. Matsuda}
\author{Yuji Aoki}\email{aoki@tmu.ac.jp}
\affiliation{%
Department of Physics, Tokyo Metropolitan University, Hachioji, Tokyo 192-0397, Japan\\
}%

\date{\today}

\begin{abstract}
Electrical resistivity, magnetic susceptibility, and specific heat measurements on single crystals of {\LaTAl} with $\it Tr\rm$ = Mo and W revealed that these compounds exhibit superconductivity with transition temperatures $\Tc$ = 3.22 and 1.81 K, respectively, achieving the highest values in the reported {\LaTAl} compounds.
There appears a positive correlation between $\Tc$ and the electronic specific heat coefficient, which increases with increasing the number of $4d$- and $5d$-electrons.
This finding indicates that filling of the upper $e_g$ orbitals in the $4d$ and $5d$ bands plays an essential role for the significant enhancement of the superconducting condensation energy.
Possible roles played by the $d$ electrons in the strongly correlated electron phenomena appearing in $\it RTr\rm_{2}$Al$_{20}$ are discussed.
\end{abstract}

\pacs{Valid PACS appear here}
\maketitle



\section{Introduction}

Ternary intermetallic compounds containing rare-earth ions are the subject of continuous interest in the fields of strongly correlated electron physics.
Among them, a family of $\it RTr\rm_{2}\it X\rm_{20}$ ($\it R\rm$ : rare earths, $\it Tr$ : transition metals, $\it X\rm$ : Al, Zn, and Cd), which crystallize in the cubic CeCr$_{2}$Al$_{20}$-type structure ($Fd\bar{3}m$, $\#227$), have attracted considerable attention in recent years, because a wide variety of exotic electron states caused by strong hybridization of $f$-electrons with non-$f$-ligands have been observed.
A heavy fermion (HF) behavior appears in YbCo$_{2}$Zn$_{20}$ with an electronic specific heat coefficient of 8 J/(mol K$^{2}$), which is the largest among Yb compounds~\cite{Torikachvili_PNAS_07, Honda_2014, Kong_2017}.
In a HF state of YbIr$_{2}$Zn$_{20}$, a metamagnetic anomaly occurs at around 10 T~\cite{Takeuchi_2010}.
Sm$\it Tr\rm_{2}$Al$_{20}$ ($\it Tr$ = Ti, V, Nb, and Ta) exhibit rare Sm-based HF behaviors, which are anomalously field-insensitive~\cite{Higashinaka_JPSJ_11_SmTi2Al20, Sakai_PRB_11, Yamada_JPSJ_13, Higashinaka_AIP_18}.
Many of Pr$\it Tr\rm_{2}\it X\rm_{20}$ compounds have a non-Kramers $\Gamma_{3}$ doublet crystalline-electric-field ground state of Pr ions, and exhibit quadrupole Kondo lattice behaviors~\cite{Cox_PRL_1987, Tsuruta_JPSJ_2015, Sakai_JPSJ_11, Onimaru_JPSJ_16, Yoshida_JPSJ_17, Higashinaka_JPSJ_17}.
Therefore, the superconductivity (SC) appearing in the Pr$\it Tr\rm_{2}\it X\rm_{20}$ compounds is presumed to be induced by quadrupolar fluctuations~\cite{Onimaru_JPSJ_11, Sakai_JPSJ_12_PrTi2Al20, Matsubayashi_PRL_12, Tsujimoto_PRL_14, Wakiya_JPSJ_17}.
In Ce- and U-based compounds, strongly correlated electron behaviors have also been reported~\cite{White_2015, Hirose_2015, Bauer_2008}.

The SC appearing in $\it RTr\rm_{2}\it X\rm_{20}$ with nonmagnetic $R$ ions has been discussed in terms of the cage structure, which is one of the characteristic features of the CeCr$_{2}$Al$_{20}$-type crystal structure.
The $\it R$ ions at the $8a$ site with cubic $\it T\rm_{d}$ symmetry are located at the center of an $\it X\rm_{16}$ cage.
In {\RxVAl} with $\it R$ = Al and Ga (the SC transition temperatures $\Tc$ are 1.49 and 1.66 K, respectively), the cage-center $R$ ions show anharmonic large-amplitude oscillations as observed in filled skutterudites~\cite{Sato}, which are considered to enhance $\Tc$ through the electron-phonon coupling~\cite{Hiroi_JPSJ_12, Onosaka_JPSJ_12, Safarik_PRB_12, Koza_PCCP_14}.
Superconductors of $\it R$ = Sc, Y, and Lu seems to have similar features~\cite{Winiarski_PRB_16}.
For recently-found superconductors of {\LaTAl} with $\it Tr\rm$ = Ti, V, Nb, and Ta ($\Tc$ ranging from 0.15 to 1.05 K)~\cite{Yamada_JPSJ_2018}, however, the cage does not have enough space for such anharmonic large-amplitude oscillations and the reason for the largely distributed $\Tc$ remains to be clarified.

In this paper, we study {\LaTAl} with $\it Tr\rm$ = Mo and W using single crystals.
The results reveal that these compounds are new superconductors with the highest $\Tc$ values among {\LaTAl} compounds.  
Comparison among all these compounds suggests that the $d$-band filling plays an essential role for determining the superconducting properties in {\LaTAl}.

\section{Experimental Details}

\begin{table}[tb]
\caption{Crystallographic parameters of La$\it Tr\rm_2$Al$_{20}$ ($\it Tr\rm =$ Mo and W) at room temperature. $R$ and $wR$ are reliability factors. $B_{\rm eq}$ is the equivalent isotropic atomic displacement parameter. Occ. is the site occupancy. Standard deviations in the positions of the least significant digits are given in parentheses.}
\label{tablestruct}
\begin{center}
\begin{tabular}{lcccclc}
\hline \hline
\multicolumn{3}{c}{LaMo$_{2}$Al$_{20}$} & \multicolumn{3}{l}{$R$ $=$ 1.94$\%$, $wR$ $=$ 3.67$\%$}\\
\multicolumn{3}{c}{$Fd\bar3m$ ($\sharp$227) (origin choice 2)} &\multicolumn{3}{l}{$a$ $=$ 14.6631(13) $\AA$, $V$ $= $ 3152.7(5) $\AA^3$}\\
\multicolumn{2}{l}{} & \multicolumn{3}{c}{Position}\\
\cline{3-5}
Atom & Site & $x$ & $y$ & $z$ & $B_{\rm eq}$($\AA^2$) & Occ.\\
\hline
La & $8a$ & 1/8 & 1/8 & 1/8 & 0.738(17) & 1 \\
Mo & $16d$ & 1/2 & 1/2 & 1/2 & 0.54(2) & 0.928(4) \\
Al(1) & $96g$ & 0.05870(5) & 0.05870(5) & 0.32549(7) & 0.82(3) & 1 \\
Al(2) & $48f$ & \ \  0.48694(10) & 1/8 & 1/8 & 0.82(3) & 1 \\
Al(3) & $16c$ & 0 & 0 & 0 & 1.79(5) & 1 \\
\hline \hline
\\
\hline \hline
\multicolumn{3}{c}{LaW$_{2}$Al$_{20}$} & \multicolumn{3}{c}{$R$ $=$ 1.04$\%$, $wR$ $=$ 2.76$\%$}\\
\multicolumn{3}{c}{$Fd\bar3m$ ($\sharp$227) (origin choice 2)} & \multicolumn{3}{c}{$a$ $=$ 14.6813(11) $\AA$, $V$ $= $ 3164.4(4) $\AA^3$}\\
\multicolumn{2}{l}{} & \multicolumn{3}{c}{Position}\\
\cline{3-5}
Atom & Site & $x$ & $y$ & $z$ & $B_{\rm eq}$($\AA^2$) & Occ.\\
\hline
La & $8a$ & 1/8 & 1/8 & 1/8 & 0.667(15) & 1 \\
W & $16d$ & 1/2 & 1/2 & 1/2 & 0.437(12) & 0.848(2) \\
Al(1) & $96g$ & 0.05872(4) & 0.05872(4) & 0.32575(6) & 0.89(2) & 1 \\
Al(2) & $48f$ & 0.48720(9) & 1/8 & 1/8 & 0.91(2) & 1 \\
Al(3) & $16c$ & 0 & 0 & 0 & 1.74(5) & 1 \\
\hline \hline
\end{tabular}
\end{center}
\end{table}

Single crystals of La$\it Tr\rm_{2}$Al$_{20}$ ($\it Tr\rm =$ Mo and W) were grown by the Al self-flux method. The starting materials were La chips (99.9$\%$), Al grains (99.99$\%$) and powders of Mo (99.99$\%$) and W (99.99$\%$). With an atomic ratio of La:Mo:Al = 1:2:50 and La:W:Al = 1:2:90, the starting materials were put in an alumina crucible and sealed in a quartz tube.
The quartz tube was heated up to 1050 $^{\rm o}$C and then slowly cooled.
Single crystals were obtained by spinning the ampoule in a centrifuge in order to remove the excess Al flux. 

The electrical resistivity $\rho$ and specific heat $C$ were measured using a Quantum Design (QD) Physical Property Measurement System (PPMS) equipped with a Helium-3 cryostat.
The magnetic susceptibility $\chi$ was measured down to 2 K using a QD Magnetic Property Measurement System (MPMS).

\section{Results and Discussion}

Single crystal X-ray diffraction analysis was performed using a Rigaku XtaLABmini with graphite monochromated Mo-K$_{\rm\alpha}$ radiation.
The structural parameters refined using the program SHELX-97~\cite{Sheldrick_SHELX-97_97} are shown in Table~\ref{tablestruct}.
The lattice parameters $a$ are close to those in the previous report~\cite{Niemann_JSSC_95}.
The equivalent isotropic atomic displacement parameter $B_{\rm eq}$ of Al(3) at the 16$c$ site has relatively large values: $B_{\rm eq}=1.74-1.79$ $\AA^2$.
This feature is characteristic to $\it RTr\rm_{2}\it X\rm_{20}$ compounds; see Refs.\cite{Nasch_ZNB_97, Kangas_JSSC_12, Yamada_JPSJ_2018} for $\it X\rm$ = Al and Refs.~\cite{Onimaru_JPSJ_11, Hasegawa_JPCS_12, Wakiya_PRB_16} for $\it X\rm$ = Zn.
The cage-center La ions at the 8$a$ site have normal values, in contrast to {\RxVAl} ($\it R$ = Al and Ga), in which the cage-center $R$ ions are suggested to have anharmonic rattling modes~\cite{Hiroi_JPSJ_12, Onosaka_JPSJ_12, Safarik_PRB_12, Koza_PCCP_14}.
The occupancy of Mo and W sites was found to be less than one. 
Similar feature was also observed for CeMo$_2$Al$_{20}$~\cite{Niemann_JSSC_95}.
This could mean that these $\it Tr\rm$ sites are partially substituted by Al atoms because of the similarity in the metallic radii~\cite{Earnshaw_97}.

\begin{figure}
\begin{center}
\includegraphics[width=0.5\linewidth]{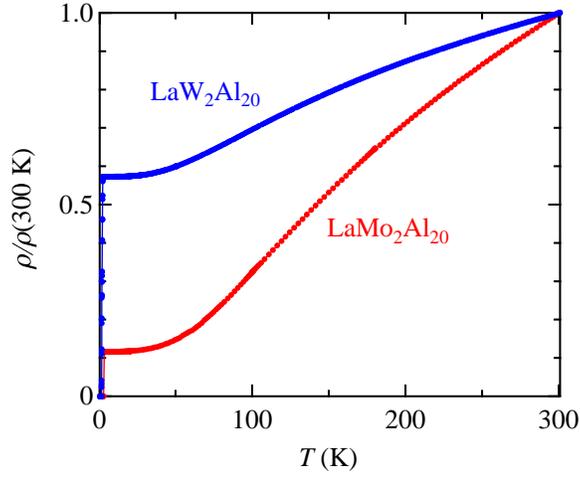}  
\end{center}
\caption{(a) Temperature dependence of electrical resistivity $\rho$ for {\LaTAl}  ($\it Tr\rm$ = Mo and W) with the current along the $\langle 110 \rangle$ direction.
}
\label{rho1}
\end{figure}

\begin{figure}
\begin{center}
\includegraphics[width=0.9\linewidth]{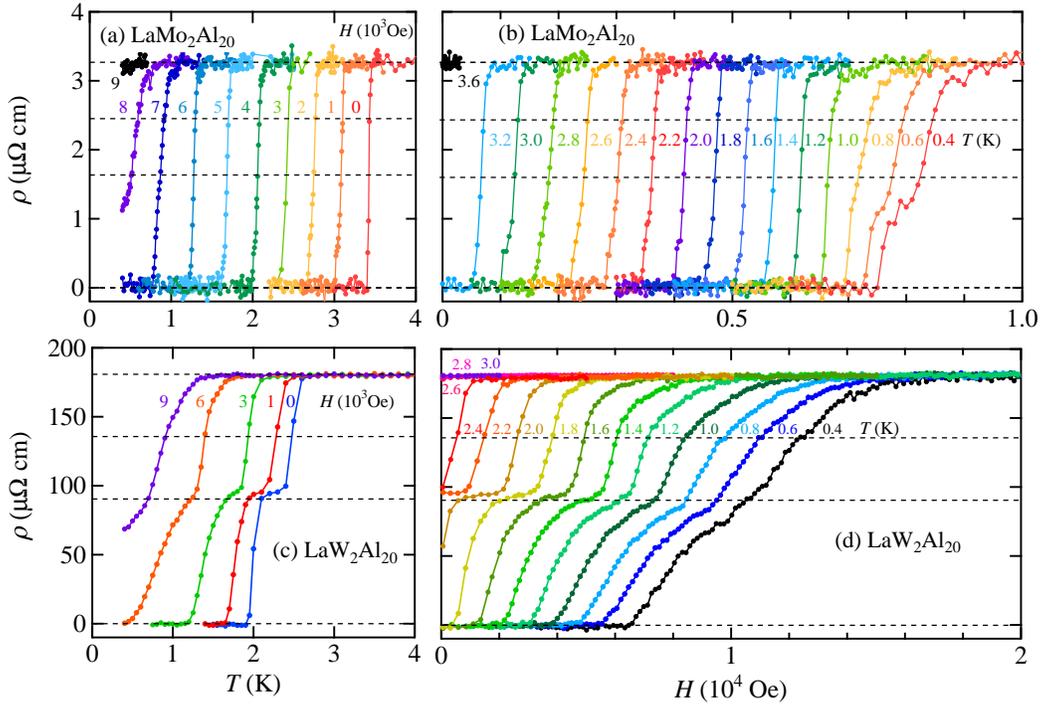}  
\end{center}
\caption{Temperature and magnetic field dependences of electrical resistivity $\rho$ for {\LaTAl} ($\it Tr\rm$ = Mo and W) measured at low temperatures with the current along the [1$\bar{1}$0] direction in the fields along the [111] direction.
}
\label{rho2}
\end{figure}

The temperature dependence of resistivity $\rho(T)$ divided by $\rho$(300 K) is shown in Fig.~\ref{rho1}. 
The residual resistivity ratio $\it RRR\rm \equiv \rho$(300 K)/$\rho_{\rm res}$ ($\rho_{\rm res}$: the residual resistivity) is 1.8 for W and 9.1 for Mo. %
Figures~\ref{rho2}(a-d) show the low-temperature expansion of $\rho(T, H)$ data.
In zero field, both compounds show SC transitions with the onset at 3.4 K for Mo and 2.6 K for W.
In the applied fields, the transition temperature shifts to lower temperatures.
The details are discussed below.

\begin{figure}
\begin{center}
\includegraphics[width=0.5\linewidth]{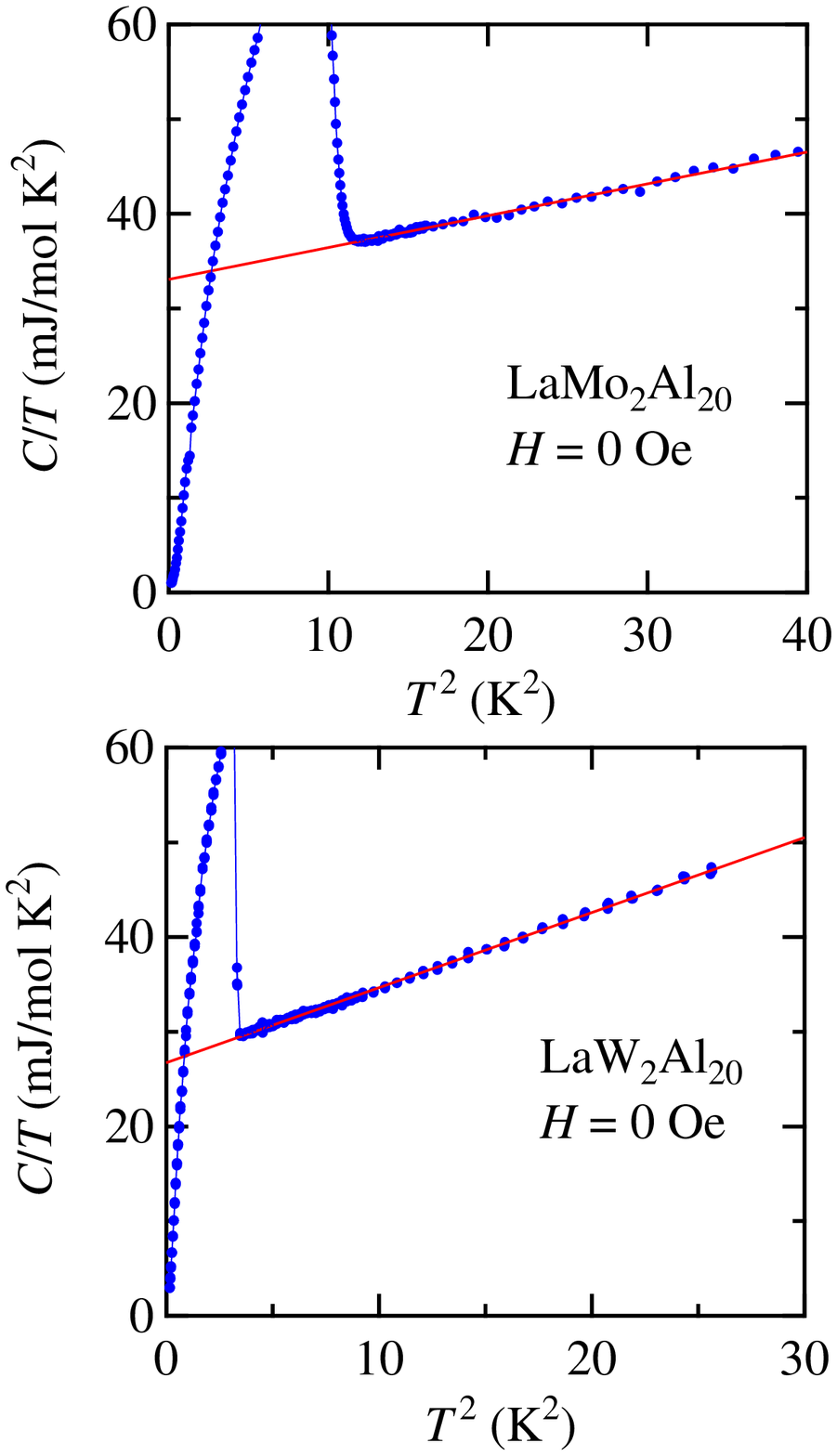}  
\end{center}
\caption{Temperature dependence of specific heat $C$ divided by temperature as a function of $T^2$ for La$\it Tr\rm_{2}$Al$_{20}$ ($\it Tr\rm$ = Mo and W).
}
\label{sh1}
\end{figure}

The temperature dependence of specific heat $C$ divided by temperature as a function of $T^2$ is shown in Fig.~\ref{sh1}.
The normal-state $C/T$ data can be well described by $C/T = \gamma + \beta T^{2}$, where $\gamma$ and $\beta$ are the electronic and phonon specific heat coefficients, respectively.
The Debye temperature $\it\Theta_{\rm D}$ is obtained from $\it\Theta_{\rm D} = \sqrt[3]{\rm(12/5)\pi^{4}\it nR/\beta}$, where $n=23$ is the number of atoms per formula unit and $R$ is the gas constant.
The obtained parameters are summarized in Table~\ref{tableSCpara}.

\begin{figure}
\begin{center}
\includegraphics[width=0.5\linewidth]{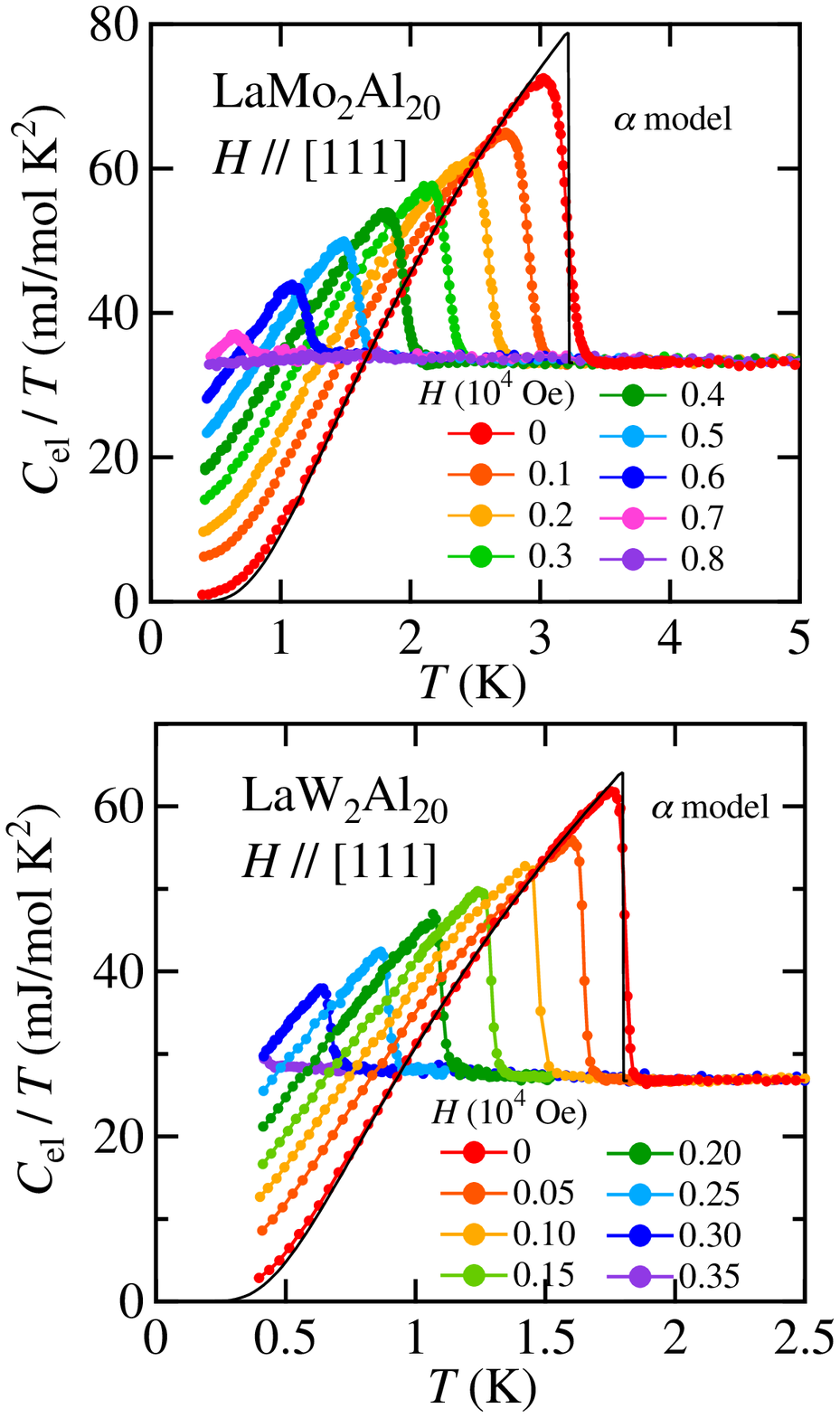}  
\end{center}
\caption{Temperature dependence of the electronic contribution to the specific heat $C_{\rm el}/T \equiv C/T-\beta T^2$ (the lower panel) for La$\it Tr\rm_{2}$Al$_{20}$ ($\it Tr\rm$ = Mo and W).
The solid curves represent the fitting by the $\alpha$ model~\cite{Johnson_SST_13, Padamsee_JLTP_73}.
}
\label{sh2}
\end{figure}

The temperature dependence of the electronic contribution to the specific heat $C_{\rm el}/T \equiv C/T-\beta T^2$ is shown in Fig.~\ref{sh2}.
A clear specific heat jump appears at 3.22 K (Mo) and 1.81 K (W), which is referred to as the bulk SC transition temperature $\Tc$ hereinafter.
The fitting of the $C_{\rm el}(T)$ data by the $\alpha$ model~\cite{Johnson_SST_13, Padamsee_JLTP_73} is shown by the solid curve. 
The obtained $\alpha$ value is 1.74 and 1.75 for Mo and W, respectively, which is close to 1.764 expected from the BCS theory, suggesting that they are weak-coupling superconductors.

\begin{figure}
\begin{center}
\includegraphics[width=0.5\linewidth]{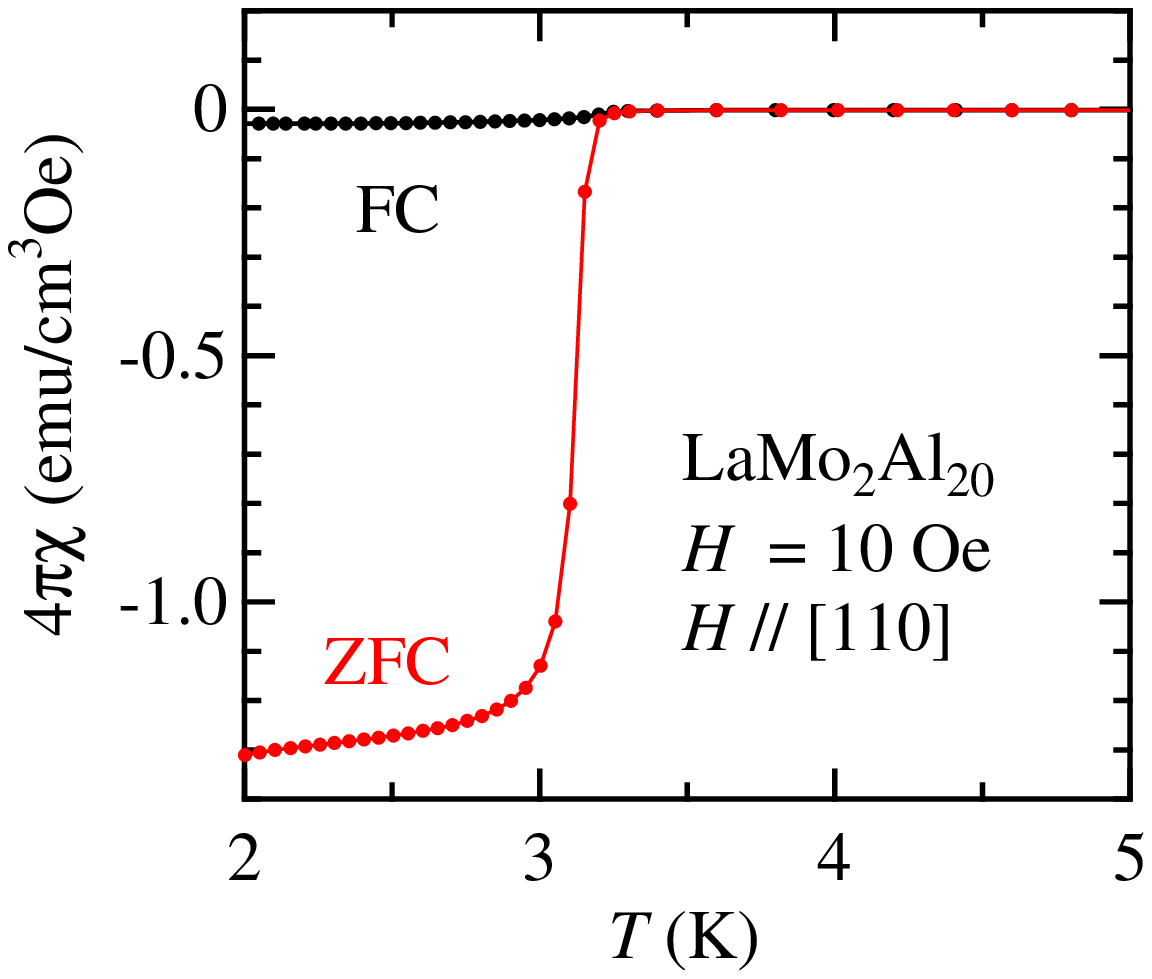}  
\end{center}
\caption{Temperature dependence of magnetic susceptibility $\chi$ for LaMo$_2$Al$_{20}$. 
The zero-field-cooled (ZFC) warming data and the field-cooled (FC) data for the applied magnetic field of 10 Oe are shown.
}
\label{chi}
\end{figure}

The bulk nature of the superconductivity in LaMo$_2$Al$_{20}$ has been confirmed by magnetic-susceptibility ($\chi$) measurements.
The temperature dependence of $\chi$ measured in 10 Oe is shown in Fig.~\ref{chi}.
The diamagnetic signal develops below $\Tc$.
The $4\pi\chi$ values of the order of -1 far below $\Tc$ suggest that the SC volume fraction reaches approximately 100$\%$.

\begin{figure}
\begin{center}
\includegraphics[width=0.5\linewidth]{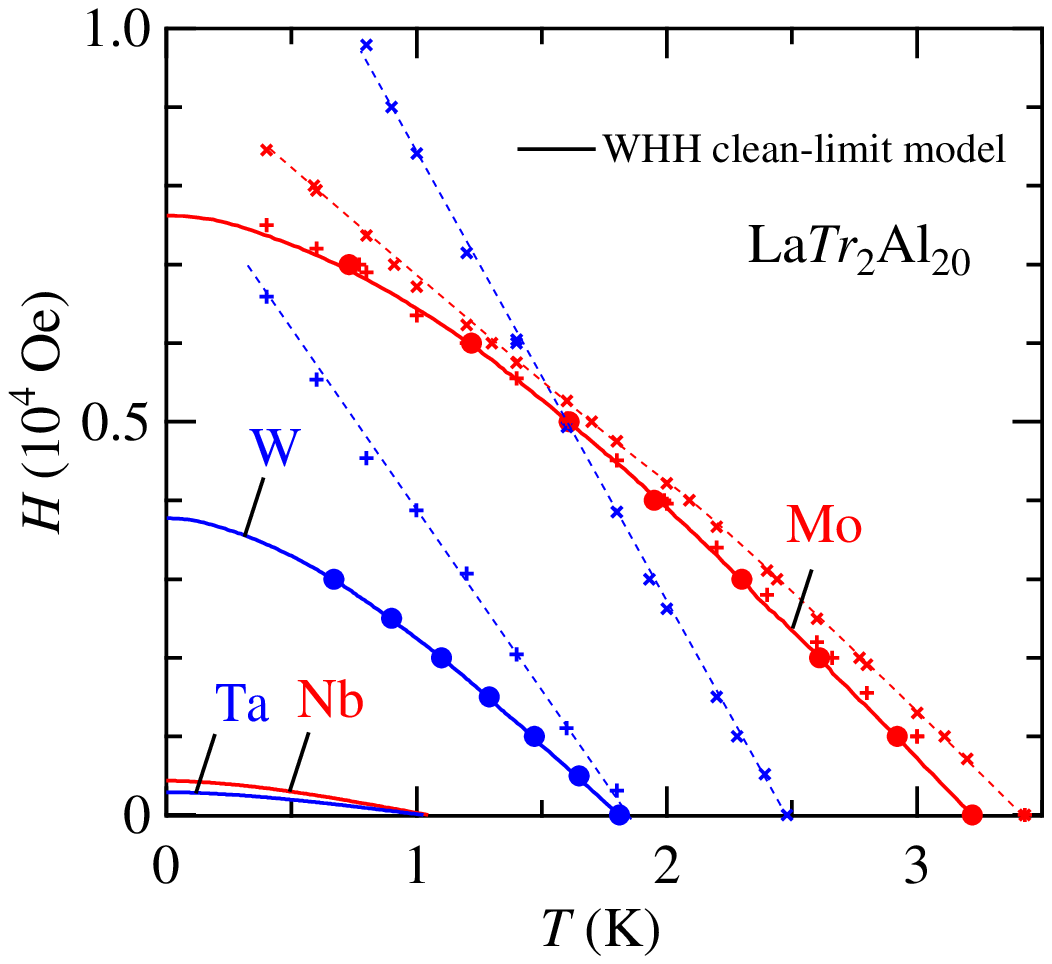}  
\end{center}
\caption{$H$-$T$ phase diagram of La$\it Tr\rm$$_{2}$Al$_{20}$ with $\it Tr\rm$ = Mo and W in comparison with $\it Tr\rm$ = Nb and Ta~\cite{Yamada_JPSJ_2018}.
Filled circles represent the bulk SC transition points obtained from the $C(T, H)$ data, which can be well described by the Werthamer-Helfand-Hohenberg (WHH) clean-limit model (solid curves)~\cite{HW_PR_66, WHH_PR_66}.
The ``cross" and ``plus" symbols designate resistive transition points defined at 75\% and 0\% of the normal-state resistance, respectively.
These points obtained from the $\rho(T, H)$ data provide higher values of $\Tc$ and $H_{\rm c2}$ compared to those from the $C(T, H)$ data, more significantly for $\it Tr\rm$ = W.
This observation indicates that a minor part of the single crystal has higher $\Tc$'s and $H_{\rm c2}$'s, which are detected by the $\rho$ measurements.
}
\label{HT}
\end{figure}

The $H$-vs-$T$ SC phase diagram constructed using the $\rho$, $C_{\rm el}$, and $\chi$ data is shown in Fig.~\ref{HT}.
The values of $H_{\rm c2}(0)$ are much lower than the Pauli-limiting field $H_{\rm P}$ = (1.84$\times 10^4$ Oe/K) $T_{\rm c}$~\cite{Clogston}, suggesting that $H_{\rm c2}(0)$ is determined by the orbital depairing effect. 
The temperature dependence of $H_{\rm c2}$ can be well described by the Werthamer-Helfand-Hohenberg (WHH) clean-limit expression~\cite{HW_PR_66, WHH_PR_66}, as shown by the solid curves in Fig.~\ref{HT}.
In this model, $H_{\rm c2}(0)$ can be expressed as
\begin{equation}
H_{\rm c2}(0) = -0.73\times\frac{dH_{\rm c2}}{dT}|_{T=T_{\rm c}}T_{\rm c} = \frac{\phi_{0}}{2 \pi \xi_{\rm GL}^2},
\label{eqWHH}
\end{equation}
where $\phi_{0}$ and $\xi_{\rm GL}$ are the quantum magnetic flux and the Ginzburg-Landau (GL) coherence length, respectively.
The GL parameter $\kappa_{\rm GL}$, which is equal to the Maki parameter~\cite{Maki} $\kappa_2(T \to T_{\rm c})$, is determined using the thermodynamic relation~\cite{Serin}: %
\begin{equation}
\frac{\Delta C_{\rm vol}}{T}\vert_{T=T_{\rm c}} = (\frac{dH_{\rm c2}}{dT}|_{T=T_{\rm c}})^2 \frac{1}{4 \pi (2 \kappa_2^2-1) \beta_{\rm A}},
\label{kappa2}
\end{equation}
where $\Delta C_{\rm vol}$ is measured per unit volume [unit: erg/(K cm$^3$)], and $\beta_{\rm A}=1.16$ for a triangular vortex lattice.
The thermodynamic critical field $H_{\rm c}(0) = \alpha \sqrt{(6/\pi) \gamma_{\rm vol}} T_{\rm c}$~\cite{Johnson_SST_13}, the London penetration depth $\lambda_{\rm L} = \kappa_{\rm GL} \xi_{\rm GL}$, and the lower critical field $H_{\rm c1} = H_{\rm c}(0) \ln \kappa_{\rm GL}/(\sqrt{2} \kappa_{\rm GL})$ are also calculated. 
The obtained characteristic parameters are summarized in Table~\ref{tableSCpara}.

The electron-phonon coupling constant $\lambda_{\rm e\textendash ph}$ is obtained using McMillan's formula
\begin{equation}
\lambda_{\rm e\textendash ph} = \frac{1.04 + \mu^{*}\ln(\frac{\it\Theta_{\rm D}}{1.45T_{\rm c}})}{(1-0.62\mu^{*})\ln(\frac{\it\Theta_{\rm D}}{1.45T_{\rm c}}) - 1.04},
\label{eqmodMcMillan}
\end{equation}
where the Coulomb coupling constant $\mu^{*}$ is assumed to be 0.13~\cite{McMillan_PR_68}.
The fact that $\lambda_{\rm e\textendash ph}=0.48-0.50$ is consistent with the above-mentioned weak-coupling nature of the superconductivity.

\begin{table*}[tb] %
\caption{Characteristic parameters of {\LaTAl} superconductors (see text for definitions).
The errors in the last significant digit(s) are indicated in parentheses.%
%
}%
\label{tableSCpara}
\begin{center}
\begin{tabular}{rcc}
\hline \hline
compounds & LaMo$_{2}$Al$_{20}$ & LaW$_{2}$Al$_{20}$\\
\hline
$\Tc$ (K) & 3.22 & 1.81\\
$\gamma$ (mJ/mol K$^{2}$) & 33.1 & 26.7\\
$\alpha$ & 1.74 & 1.75 \\
$\Delta C/\gamma\Tc$ & 1.32 & 1.37\\
$\it\Theta_{\rm D}$ (K) & 511 & 383\\
$\lambda_{\rm e\textendash ph}$ & 0.504 & 0.476\\
$H_{\rm c}$(0) (Oe)  & 219 & 111 \\
$\frac{dH_{\rm c2}}{dT}\vert_{T=T_{\rm c}}$ (Oe/K) & -3240 & -2860 \\ 
$H_{\rm c2}$(0) (Oe)  & 7620 & 3770\\ 
$\xi_{\rm GL}$ ($\AA$) & 207 & 295 \\ 
$\kappa_{\rm GL}=\kappa_2(T \to T_{\rm c})$ & 14.0 & 13.5 \\ 
$\lambda_{\rm L}=\kappa_{\rm GL} \xi_{\rm GL}$ ($\AA$) & 2900 & 3980 \\  
$H_{\rm c1}$(0) (Oe)  & 29 & 15 \\ 
\hline \hline
\end{tabular}
\end{center}
\end{table*}

\begin{figure}
\begin{center}
\includegraphics[width=0.5\linewidth]{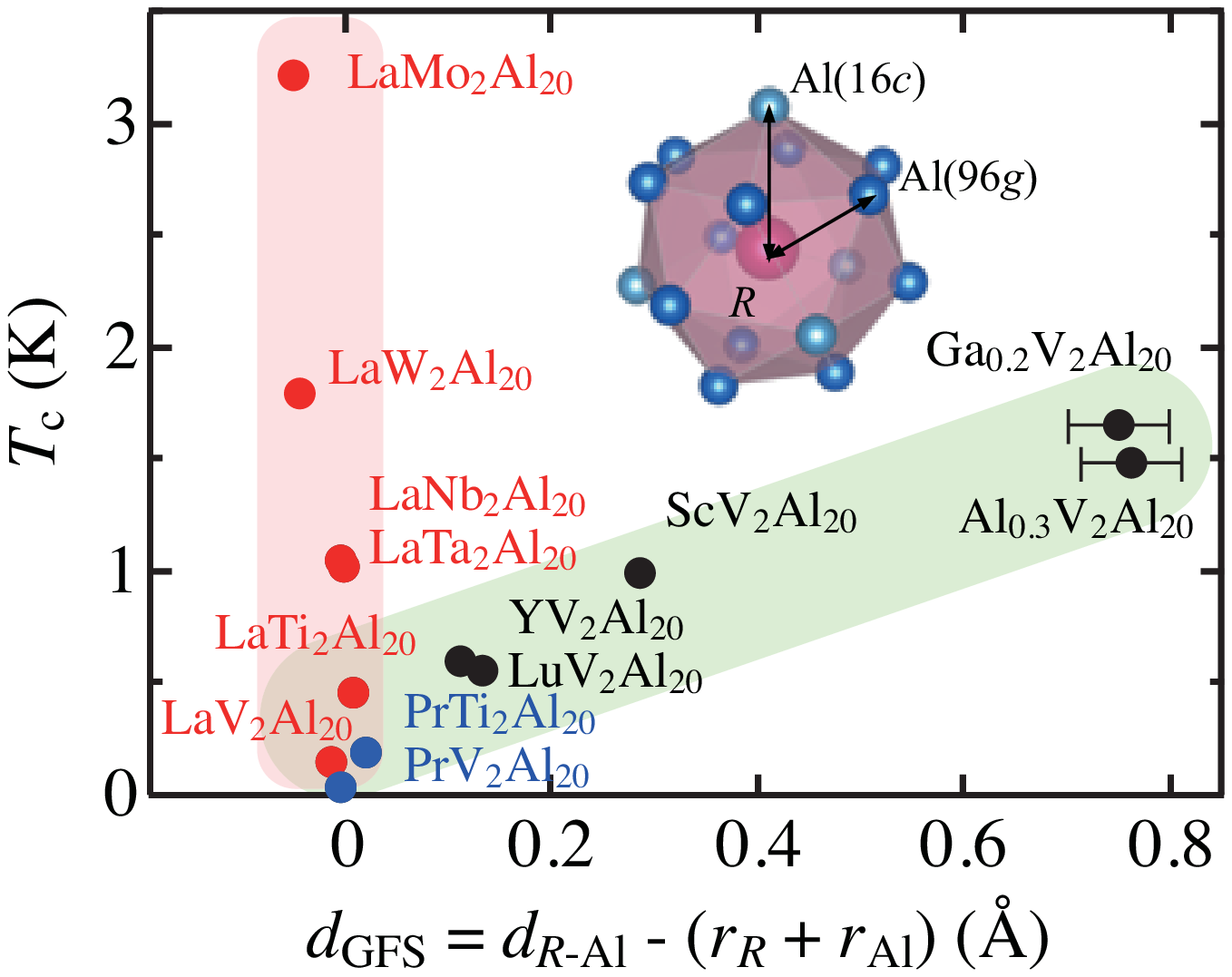}  
\end{center}
\caption{$\Tc$ vs $d_{\rm GFS} \equiv \it d_{R \rm \--Al}-(\it r_{R}+\it r_{\rm Al})$ quantifying the ``guest free space" of nonmagnetic cage-center $R$ ions for $\it RTr_{2}$Al$_{20}$ (see text for details).
The data except for LaMo$_2$Al$_{20}$ and LaW$_2$Al$_{20}$ are taken from Fig.5 in Ref.~\cite{Yamada_JPSJ_2018}. 
This figure demonstrates that nonmagnetic $\it RTr_{2}$Al$_{20}$ superconductors are classified into two groups, i.e., (A) $d_{\rm GFS} \ne 0$ and $\Tc$ correlates with $d_{\rm GFS}$, and (B) $d_{\rm GFS} \simeq 0$ and $\Tc$ seems to be governed by other factors.
Note that superconductors PrTi$_2$Al$_{20}$~\cite{Sakai_JPSJ_12_PrTi2Al20} and PrV$_2$Al$_{20}$~\cite{Tsujimoto_PRL_14}, and field-insensitive HF compounds Sm$\it Tr_2$Al$_{20}$ ($\it Tr$= Ti, V, Cr, and Ta)~\cite{Higashinaka_JPSJ_11_SmTi2Al20, Sakai_PRB_11, Yamada_JPSJ_13} also have $d_{\rm GFS} \simeq 0$. %
The inset picture shows the structure of a $R(8a)$-Al$_{16}(96g, 16c)$ cage; Al$_{16}$ forms a CN 16 Frank-Kasper polyhedron.
}
\label{Tcgfs}
\end{figure}

In the crystal structure of $\it RTr\rm_{2}$Al$_{20}$, the Al$_{16}$ cage includes a guest $R$ ion at the center ($8a$ site) as shown in the inset of Fig.~\ref{Tcgfs}.
In Ref.~\cite{Yamada_JPSJ_2018}, we have introduced a parameter to quantify the ``guest free space" as $d_{\rm GFS} \equiv d_{R \rm \--Al}-(\it r_{R}+\it r_{\rm Al})$, where $d_{R \rm \--Al} \equiv (12 d_{R \rm \--Al(96g)}+4 d_{R \rm \--Al(16c)})/16$ is the average distance between $R$ and Al in the cage, and $r_{R}$ and $r_{\rm Al}$ are the covalent radii for $\it R$ and Al ions, respectively~\cite{Cordero_DT_08}.
$d_{R \rm \--Al}$ is calculated using the results of the single-crystal X-ray diffraction analysis shown in Table~\ref{tablestruct}.
In Fig.~\ref{Tcgfs}, we show $\Tc$ vs. $\it d_{\rm GFS}$ for $\it RTr_{2}$Al$_{20}$ with nonmagnetic $\it R$ ions; this is a revised one of Fig. 5 in Ref.~\cite{Yamada_JPSJ_2018}.
Nonmagnetic $\it RTr_{2}$Al$_{20}$ superconductors are classified into two groups, i.e., (A) $d_{\rm GFS} \ne 0$ and $\Tc$ correlates with $d_{\rm GFS}$, and (B) $d_{\rm GFS} \simeq 0$ and $\Tc$ seems to be governed by other factors.
For group (A), it is thought that $\Tc$ is enhanced by the ``rattling" anharmonic vibration modes of Ga, Al, Sc, and Lu ions due to the coupling with conduction electrons~\cite{Hiroi_JPSJ_12, Onosaka_JPSJ_12, Safarik_PRB_12, Koza_PCCP_14, Winiarski_PRB_16}.
In contrast, all the data points in group (B) fall almost into a vertical line with $\it d_{\rm GFS} \rm\simeq 0$, indicating that these {\LaTAl} compounds do not have guest free space and the large $\Tc$ distribution is not associated with the La ion oscillations.

\begin{figure}
\begin{center}
\includegraphics[width=0.5\linewidth]{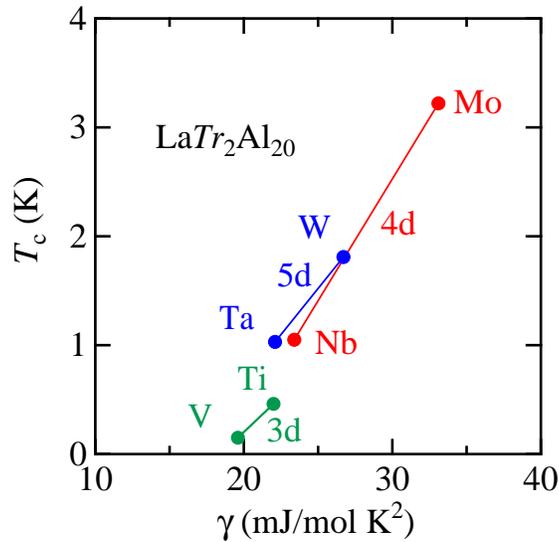}   
\end{center}
\caption{$\Tc$ vs the electronic specific heat coefficient $\gamma$ for {\LaTAl}.
}
\label{Tcgamma}
\end{figure}

%

The distribution of $\Tc$ among {\LaTAl} is remarkably large; $\Tc(\it Tr=$Mo$)/\Tc(\it Tr=$V$)=3.22/0.15 \simeq 22$.
Figure~\ref{Tcgamma} shows $\Tc$ vs. the electronic specific heat coefficient $\gamma$ for all {\LaTAl} superconductors.
This figure clearly demonstrates that there is a positive correlation between $\Tc$ and $\gamma$.
$\it Tr\rm$ ions are located at sites with trigonal point symmetry $D_{3d}$.
Due to the crystalline-electric field effect, the fivefold degenerate $d$ orbitals of a $\it Tr\rm$ ion split into a low-energy singlet ($a_{1g}$) and two high-energy doublets ($e_g$)~\cite{Swatek_2018}.
Electronic band structure calculations for $\it Tr\rm$ = Ti, V, and Cr~\cite{Swatek_2018} suggest that there is a ferromagnetic instability, which becomes more dominant with $3d$-electron filling into the upper $e_g$ orbitals approaching Cr; the calculated Stoner factor of LaCr$_2$Al$_{20}$ is relatively high although no ferromagnetic ordering has been observed experimentally.
This instability may be one of the possible reasons for the suppressed $\Tc$ values for those $3d$ compounds.
On the contrary, for the $4d$ and $5d$ compounds, Fig.~\ref{Tcgamma} demonstrates that the $4d$($5d$) electron filling with Nb$\to$Mo (Ta$\to$W) boosts up the $\Tc$ value.
The increased $\gamma$ values with the electron filling indicate enhancements in the density of states at the Fermi energy and/or in the effective mass of conduction electrons for $\it Tr\rm$ = Mo and W.
Actually, as shown in Fig.~\ref{HT}, $dH_{\rm c2}/dT \vert_{T=T_{\rm c}}$ increases as Nb$\to$Mo (Ta$\to$W), providing evidence for the mass enhancement.
With these features, we speculate that the filling of the upper $e_g$ orbitals in the $4d$($5d$) bands significantly enhances the SC condensation energy.

\section{Summary}
We have studied the electrical resistivity, magnetic susceptibility, and specific heat of single crystalline LaMo$_2$Al$_{20}$ and LaW$_2$Al$_{20}$.
It has been revealed that these compounds exhibit superconductivity with transition temperatures $\Tc$ = 3.22 and 1.81 K, respectively, achieving the highest values in the reported {\LaTAl} compounds.
The values of $\Tc$ exhibit a positive correlation with the electronic specific heat coefficient $\gamma$, which increases with the $4d$ and $5d$ electron filling. 
This finding indicates that the upper $e_g$ orbitals in the $4d$ and $5d$ bands play an essential role for the significant enhancement of the SC condensation energy.

In the realization of the several types of strongly correlated electron phenomena in $\it RTr\rm_{2}$Al$_{20}$, the roles played by $d$ electrons have not been clarified yet.
According to the calculated band structures~\cite{Swatek_2018}, the Fermi surface structures change drastically with the $d$ band filling.
Therefore, the strength of hybridization with the $f$ electrons of $R$ ions is expected to change depending on the $\it Tr$ elements.
Further studies on the features of $d$ electron orbitals in $\it RTr\rm_{2}$Al$_{20}$ may help to understand the unsolved problems in Sm-based field-insensitive heavy-fermion behaviors and Pr-based quadrupole Kondo lattice behaviors accompanied by superconductivity induced by quadrupolar fluctuations.

\begin{acknowledgments}

This work was supported by MEXT/JSPS KAKENHI Grant Numbers 15H03693, 15H05884, 15J07600, 15K05178, 19H01839 and Tokyo Metropolitan Government Advanced Research Grant Number (H31-1).

\end{acknowledgments}

\nocite{*}

\bibliography{apssamp}

\end{document}